\documentclass{article}

\usepackage{arxiv}

\usepackage[utf8]{inputenc} 
\usepackage[T1]{fontenc}    
\usepackage{hyperref}       
\usepackage{url}            
\usepackage{booktabs}       
\usepackage{amsfonts}       
\usepackage{nicefrac}       
\usepackage{microtype}      
\usepackage{amsmath}
\usepackage{lipsum}
\usepackage{authblk}
\usepackage{lmodern}
\usepackage{graphicx}
\graphicspath{ {./Regional_Thermodynamic_Trends_of_AI_Emulators/} }

\title{Benchmarking Regional Thermodynamic Trends in an AI emulator, ACE2, and a hybrid model, NeuralGCM}

\author[1]{\textbf{K. Rucker\thanks{Corresponding author: krucker01@uchicago.edu}}}
\author[1]{\textbf{I. Baxter}}
\author[1]{\textbf{P. Hassanzadeh}}
\author[1]{\textbf{T. A. Shaw}}
\author[1,2]{\textbf{H. A. Pahlavan}}

\affil[1]{Department of the Geophysical Sciences, The University of Chicago, Chicago, IL}
\affil[2]{Northwest Research Associates, Boulder, CO}

\begin{document}
\maketitle



%
%

%
%

\begin{abstract}
 AI models have emerged as potential complements to physics-based models, but their skill in capturing observed regional climate trends with important societal impacts has not been explored.  Here, we benchmark satellite-era regional thermodynamic trends, including extremes, in an AI emulator (ACE2) and a hybrid model (NeuralGCM). We also compare the AI models' skill to physics-based land-atmosphere models.  Both AI models show skill in capturing regional temperature trends such as Arctic Amplification. ACE2 outperforms other models in capturing vertical temperature trends in the midlatitudes. However, the AI models do not capture regional trends in heat extremes over the US Southwest. Furthermore, they do not capture drying trends in arid regions, even though they generally perform better than physics-based models. Our results also show that a data-driven AI emulator can perform comparably to, or better than, hybrid and physics-based models in capturing regional thermodynamic trends.
\end{abstract}


\section*{Keypoints}
\begin{itemize}
    \item AI models capture Arctic Amplification and outperform physics-based models in capturing vertical tropical temperature trends.
    \item Fully data-driven ACE2 outperforms other models in capturing the vertical structure of extratropical warming trends.
    \item Neither AI nor physics-based models capture extreme heat or drying trends over the Southwest US.
\end{itemize}

\section*{Plain Language Summary}
Predicting changes in temperature and humidity on the regional scale is important for climate adaptation and risk mitigation efforts.  Up until recently, physics-based climate models have been the primary tool used to predict the effects of climate change. In the past few years, AI emulators and hybrid models, which use neural networks trained on global observational products, have emerged. These AI models are optimized for, and perform well on, global metrics, but their skill and trustworthiness have not been fully tested at regional scales.  Here, we examine two such models in their ability to capture regional trends in mean temperature, heat extremes, and drying. We show that AI models exhibit better skill in predicting trends in some regions where physics-based models struggle, for example, tropical warming aloft. However, we find examples where neither AI nor physics-based models can capture recent regional changes, for example, the drying over the Southwest US.  These findings show AI models have great potential but further improvements are needed for skillful prediction of certain regional climate changes.

%
%

\section{Introduction}

Many regional climate change signals have already emerged from the noise in observations, including for extreme events \cite{IPCC2021}. To date, physics-based climate models have been the primary tool used to predict regional changes.  Physics-based models have successfully predicted regional trends such as stronger warming over land compared to the ocean \cite{Wallace_Joshi2018} and Arctic Amplification \cite{Hahn2021}. The latter prediction was made decades before it was observed \cite{Manabe&Wetherald}. In addition to these successes, regional discrepancies have also emerged \cite{Shaw2024}. For example, in the tropics, physics-based models overestimate the warming of the tropical upper-troposphere in the satellite era \cite{Vert_profile_CMIP6}. Other concerning discrepancies include the failure to capture trends in heat extremes over the Midwest US and Western Europe \cite{Singh2023} and drying trends in arid and semi-arid regions \cite{Simpson2023}. These discrepancies are cases where the observed signals lie outside the distribution of trends from model ensembles that capture internal variability and structural model uncertainty. It has been argued that these discrepancies may be tied to the representation of subgrid-scale physics, such as convection and land-atmosphere interactions in climate models \cite{shaw2025}.

Building on the recent success of AI weather models \cite{forecastnet, pangu-weather, graphcast, Ben_Bouall_gue_2024}, AI atmosphere emulators and hybrid models that are stable on multi-decadal timescales have emerged as promising complements to physics-based models \cite{ace2, ngcm, camulator, DLESym, SamudrACE, LUCIE}. These models build on the methodology of AI weather models, training on fast timescale processes, e.g., using 6-hourly predictions in the loss function. Most notably, \cite{ace2} and \cite{ngcm} introduced models that are trained on ERA5 reanalysis data and forced by sea-surface temperature (SST) and sea-ice concentration (SIC) over the satellite era. They have shown skill in reproducing global warming trends, tropical cyclone activity, and precipitation climatology \cite{Mu-ting2025, PNNL_Hierarchical, ace2, ngcm}. However, the skill of AI models in capturing regional thermodynamic trends over the satellite era remains untested.

Benchmarking the ability of AI models in capturing {\it observed} regional multi-decadal trends is an important test of their ability to perform a climate task. Regional climate trends, which represent the forced response in most cases, are a fundamentally different benchmark as compared to those based on out-of-distribution responses such as imposed 2K to 4K SST warming, which compare AI models to physics-based models \cite{+2k_emulation, PNNL_Hierarchical, camulator}, and unforced atmospheric variability \cite{Benchmark_variability, PNNL_Hierarchical}.  Because AI models are trained on an ERA5 weather task, i.e. short-term trajectory (fast timescale), there is no guarantee that this will lead to skill on the climate task, e.g., regional trends over the satellite era.  Furthermore, AI models are trained with a global loss function, which does not ensure success at regional scales. Also, AI models do not include land features, which, for example, could cause the models to struggle with extremes over land \cite{ngcm_heatwaves_land_feedbacks}. Furthermore, the blurring from spectral bias~\cite{chattopadhyay2023challenges,Bonavita_2024,Lai_2025} and data imbalance, i.e., a data-driven model's inability to learn infrequent events \cite{Sun_2025,sun2025predicting, Materia_ai_for_climate_extremes}, may prevent models from capturing the magnitude of extremes. 

Additionally, ``the best'' AI model is often selected from an array of hyperparameters, random seeds, and checkpoints to find the version that performs best on some metrics on climate timescales, e.g., showing minimum global time-mean bias \cite{ace2}.  Therefore, skill on the regional climate task, such as thermodynamic trends, is not guaranteed. Quantifying their performance on in-distribution regional trends offers the opportunity to test them against {\it observed} signals and would give insights into their skill on a climate task.

Here we benchmark regional thermodynamic trends, including mean and extreme temperature and drying, in the AI2 Climate Emulator 2 (ACE2), a data-driven deep learning neural network-based emulator \cite{ace2}, and  Neural General Circulation Model (NeuralGCM), a hybrid model that has a dynamical core and a neural network-based parameterization \cite{ngcm}. We choose these models because they are trained on ERA5 and they can be run under the Atmospheric Model Intercomparison Project (AMIP) protocol \cite{gates1992} developed for physics-based models. 

We compare ACE2 and NeuralGCM to physics-based land-atmosphere (AMIP) models and ERA5 reanalysis data. We focus on a few regional trends that have received considerable attention in the literature: the latitudinal temperature trends, including Arctic Amplification and upper-tropospheric warming, regional trends in heat extremes over Western Europe and the Midwest US, and drying trends over the Southwest US and other arid regions. In what follows, we describe the AI and physics-based models. Then we present their skill in capturing thermodynamic trends in the satellite era.  Finally, we summarize our conclusions and discuss implications for using AI models to improve regional climate predictions.

\section{Models \& Diagnostics}

In this section, we describe the physics-based, AI emulator, and hybrid model ensembles as well as the diagnostics used in this study.  

\subsection{Models}

\subsubsection{Physics-Based Model Ensemble}

AMIP is a protocol used to assess the atmospheric component of physics-based general circulation models \cite{gates1992}.  The models are run with prescribed SST and SIC from the Merged Hadley-NOAA/OI dataset from 1979 to 2014 to remove the complexity of atmosphere-ocean interactions.  The land surface is not prescribed and is therefore predicted as part of the simulation. Well-mixed greenhouse gases and aerosols follow CMIP6 forcings \cite{eyring_CMIP6}. We use AMIP data from the CMIP6 archive (see Table~S1 for a list of the models).  We use all available ensemble members that provided data needed for the trend analysis.  In our comparison, this means that ensembles could range from 30 to 100 members depending on data availability (Table~S1). The AMIP ensemble captures both structural and internal variability uncertainties. In order to quantify uncertainty only due to internal variability, we also quantify trends in the CAM6 ensemble, which has 10 members. In what follows, the AMIP model trend distribution is said to capture the ERA5 trend when the ERA5 trend is within the 5-95\% of the AMIP model trend distribution. 

\subsubsection{AI Climate Emulator 2 (ACE2) Ensemble}

ACE2 is a fully data-driven model that is trained with a global mean squared error loss:
\begin{equation}
\label{eq:loss}
\mathcal{L}= \frac{1}{N}\sum^N_{i=1} \|x_{i}(t+\Delta t) - \mathcal{N}(x_{i}(t),b_i(t),\theta)\|_2^2.
\end{equation}
\noindent Here, $\mathcal{N}$ is a deep neural network based on the spherical Fourier neural operator (SFNO) architecture, $x$ is the atmospheric state evolved at time intervals of $\Delta t = 6$ hours, $b(t)$ represents boundary conditions and forcings, $\theta$ are the trainable parameters of $\mathcal{N}$, and $N$ denotes the number of training samples. We use ACE2-ERA5, which is trained on $1^{\circ}$ ERA5 data from 1940-1995, 2011-2019, and 2021-2022, and validated on 1996-2000. SST, SIC, global mean atmospheric carbon dioxide, and downward shortwave radiative flux at the top of the atmosphere, which constitute $b(t)$, are prescribed from the ERA5 data provided in the ACE2 repository. The model has no explicit land component but includes skin temperature of land as a prognostic variable (in $x$). ACE2-ERA5 does not include aerosol forcings. We run 37 ensemble members, each initialized at 10-day intervals throughout 1980 and integrated through 2022. ACE2 is trained on ERA5 data vertically aggregated into eight hybrid sigma–pressure (terrain-following) layers and produces output on the same levels. Therefore, all ACE2 data have been interpolated to pressure levels using the NCAR Command Language vinth2p ECMWF function (See Fig. S1). As for AMIP, the ACE2's trend distribution is said to capture the ERA5 trend when the ERA5 trend is within the 5-95\% of the ACE2's trend distribution. 

\subsubsection{Neural General Circulation Model (NeuralGCM) Ensemble}

NeuralGCM is a hybrid model that uses a dynamical core to simulate large-scale atmosphere dynamics and a learned neural network to represent the unresolved subgrid-scale tendencies \cite{ngcm}.  The neural network has been trained on the residual between the output of the dynamical core and the ERA5 dataset during 1979-2017 on 6-hour up to 5-day rollouts, with a loss function similar to Eq.~(\ref{eq:loss}), and validated on 2018. SST, SIC, and incident solar radiation are prescribed from the ERA5 dataset. There is no specific land component, though the model does include a surface embedding that takes the lowest level atmospheric temperature and moisture as inputs. NeuralGCM does not include any greenhouse gas or aerosol forcings. We run 37 ensemble members initialized in the same way as described for ACE2 and run until 2023. We use the deterministic model with 2.8$^{\circ}$ horizontal resolution, which outputs data mapped to 37 vertical pressure levels in a decoder step.  Because the model is trained on 32 equidistant sigma levels derived from ERA5's 37 pressure levels, the loss is only minimized up to 0.0032 sigma. To maintain simulation stability, the global mean surface pressure is fixed following \cite{ngcm-imerg}.  As for AMIP, the NeuralGCM's trend distribution is said to capture the ERA5 trend when the ERA5 trend is within the 5-95\% of the NeuralGCM's trend distribution. 

\subsection{Diagnostics}

In this study, temperature and humidity are averaged on a yearly and regional basis using a latitude-weighted mean: 
\begin{equation}
    \frac{\frac{1}{12}\sum_{n=1}^{12}\sum_{i,j}\cos(\phi_i)(X(m_n,l,\phi_i,\lambda_j,e)\cdot M(l,\phi_i,\lambda_j))}{\sum_{i,j}\lambda_j\cos(\phi_i)} = \overline{X}(y,l,e)
\end{equation}
\noindent where $\Sigma_{i,j}$ is the spatial averaging operator defined over a limited longitude, $\lambda_i$, and latitude, $\phi_j$, region, $\sum_n^{12}$ is the annual averaging operator, $X$ is the thermodynamic variable of interest, $M$ is a surface pressure mask for topography, $m_n$ is the month of the year, $l$ is the pressure level, and $e$ is ensemble member.  A linear regression is performed on $\overline{X}(y,l,e)$ for the time period of interest. Trends are calculated over 1981-2014, which is the overlapping time for the different models. The emulations start in 1981 (to allow for the comparison of 37 ensemble members) and 2014 is the last year of the AMIP simulations.

We quantify thermodynamic trends for mean temperature over tropical (20$^{\circ}$S-20$^{\circ}$N), midlatitude (20$^{\circ}$-60$^{\circ}$) and polar (60$^{\circ}$-90$^{\circ}$) regions. Extreme temperature trends are calculated over Western Europe and the United States following previous work \cite{Singh2023,Simpson2023}. Because ACE2 and NeuralGCM outputs are on 6-hour intervals, daily maximum temperature is selected from the four daily samples. The maximum of these daily maxima over each year is then defined as TXx. For consistency, the same is done for ERA5 data, however, the daily maximum temperature variable is used for AMIP ensemble extreme temperature analysis.  For drying trends we focus on arid and semi-arid regions previously defined by \cite{Simpson2023}. The time series for all variables is normalized by the 1981-1990 average following the previous work of \cite{Simpson2023} and the beginning of the emulations. For regions without significant topography, we compare 2-meter trends of physics-based models and ACE2 with 1000-hPa trends of NeuralGCM because NeuralGCM does not output variables at 2-meter height.

ACE2 directly outputs temperature at 850-hPa which is used for analysis in Figure 1, otherwise all other ACE2 outputs on pressure levels that are displayed have been interpolated from its eight hybrid-sigma levels. We consider that models capture trends if ERA5 falls within the 5-95\% range of the ensemble distribution.  If ERA5 falls below this range, the models overestimate the trend, and if ERA5 falls above, the models underestimate it.

\section{Results} 

\begin{figure}
\includegraphics{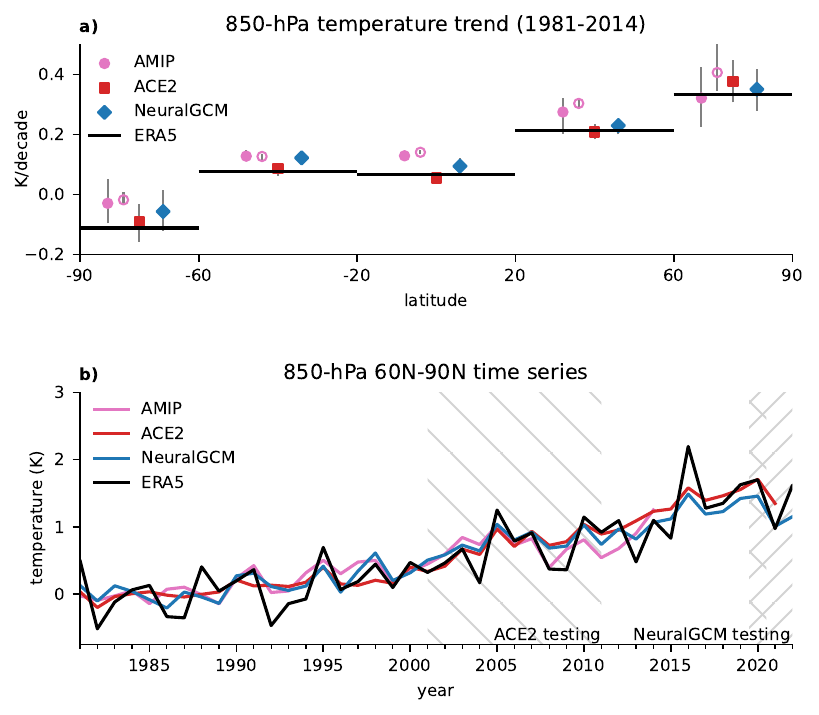}
\caption{(a) Latitude-weighted zonal-mean 850-hPa temperature trend (1981-2014) in different latitudinal bands for physics-based models, NeuralGCM, ACE2, and ERA5.  Open pink circle represents the CAM6 ensemble. Grey vertical lines represent the 5-95\% ensemble spread. Black horizontal line is the ERA5 trend. (b) Ensemble-mean annual time series of latitude-weighted 850-hPa temperature averaged over 60N-90N, normalized by the 1981-1990 mean of the respective model. Backslash hatching represents the ACE2 testing period, and forward-slash hatching represents the NeuralGCM testing period. See Figs. S2 and S3 for the same analysis but over land and over ocean, respectively.}
\label{Fig. 1}
\end{figure}

\subsection{Temperature Trends across Latitude Bands}

As mentioned in the Introduction, two key thermodynamic trends over the satellite era are Arctic Amplification \cite{Hahn2021} and stronger warming over land compared to ocean, implying an extratropical hemispheric warming contrast \cite{Wallace_Joshi2018}. Physics-based models capture near-surface (850-hPa) Arctic Amplification (Fig.~\ref{Fig. 1}a, b) consistent with the prescribed change in SIC and SST in the models. The physics-based model ensemble also captures near-surface warming of the Northern extratropics at the lower end of the ensemble distribution of trends. However, they overestimate near-surface warming in the Southern extratropics and the tropics (Fig.~\ref{Fig. 1}a). This is due to an overestimation of warming over the ocean (See Figs. S2 and S3), which is surprising given that SSTs are prescribed in the models. The CAM6 ensemble trends, which quantify uncertainty due to internal variability, agree with the entire AMIP ensemble outside of the Arctic, but overestimate Arctic Amplification (Fig.~\ref{Fig. 1}a). 

The ACE2 and NeuralGCM model ensembles perform as well or better than the physics-based model ensemble in capturing Arctic Amplification (Fig.~\ref{Fig. 1}a). Their performance is also comparable to physics-based models over time, and they show consistent performance between their testing and training periods (Fig.~\ref{Fig. 1}b). ACE2 and NeuralGCM's ability to capture Arctic Amplification is once again likely related to the SIC and SST being imposed. ACE2 performs best in capturing near-surface warming trends across the tropics and extratropics (Fig.~\ref{Fig. 1}a).  NeuralGCM overestimates the ERA5 near-surface warming trends in the Southern midlatitudes and tropics, but captures warming in the Northern extratropics. 

The skill of ACE2 and NeuralGCM in capturing some latitudinal near-surface temperature trends is noteworthy given they are trained to minimize global-mean short-term prediction error~(Eq.~(\ref{eq:loss})). ACE2's ability to capture near-surface temperature trends and outperform NeuralGCM is surprising given that NeuralGCM has a dynamical core and ACE2 is fully data-driven. 

\begin{figure}
\includegraphics{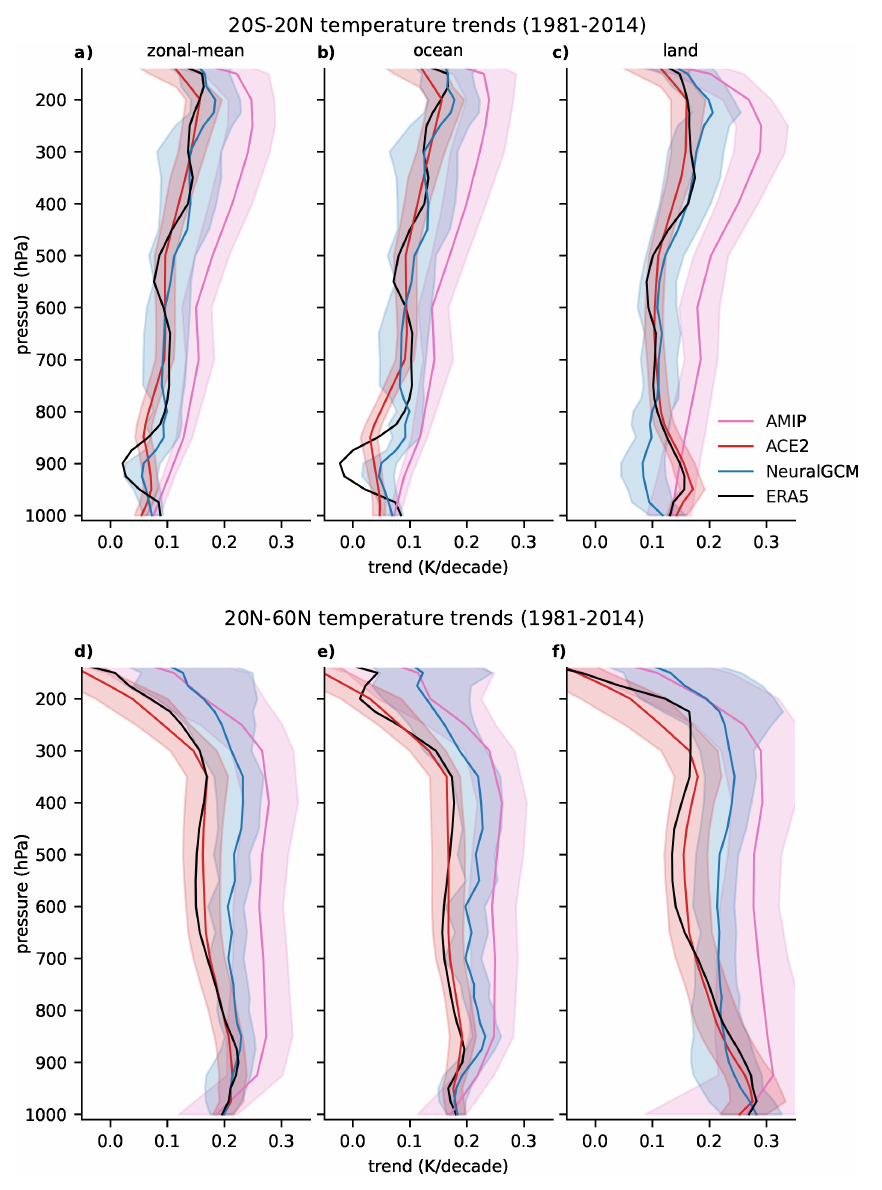}
\caption{(a)-(c) Tropical temperature trends: Temporal (1981-2014) and spatial (20S-20N) average of vertical temperature trends from physics-based models, ACE2, and NeuralGCM compared with ERA5.  Shading shows the 5-95\% ensemble spread. (d)-(f) Same as (a)-(c) but for the Northern Hemisphere extratropics (20N-60N). See Fig.~S4 for the Southern Hemisphere extratropics.}
\label{Fig. 2}
\end{figure}

\subsection{Vertical temperature trends}

In the tropics, the warming trend maximizes aloft consistent with early climate model predictions \cite{Manabe&Wetherald}. However, previous work reported physics-based models tend to simulate stronger warming in the tropical upper troposphere as compared to observations, though there are sensitivities to the imposed SST and internal variability \cite{Po-Chedley2022,Fueglistaler2025}. The overestimated warming has been connected to the representation of convection in these models, including the role of entrainment \cite{Mitchell_2013}. 

Consistent with previous studies, we also find that the observed upper tropospheric warming trend lies outside the 5-95\% of the physics-based model ensemble trend distribution (Fig.~\ref{Fig. 2}a). Furthermore, the ERA5 trend falls outside the 5-95\% ensemble range above 975-hPa. Much of the tropical mean signal reflects the trend over the ocean where physics-based models overestimate warming throughout the atmosphere (Fig.~\ref{Fig. 2}b).  Over land, physics-based models capture near-surface warming, but diverge from ERA5 aloft (Fig.~\ref{Fig. 2}c). 

NeuralGCM was previously shown to capture tropical upper tropospheric warming in ERA5 \cite{ngcm}. Here we find both ACE2 and NeuralGCM outperform physics-based models in terms of tropical temperature trends. Both AI models capture the warming aloft in the tropics (Fig.~\ref{Fig. 2}a) regardless of their training or testing period (See Fig. S5). However, these AI models do not show similar performance in capturing tropical trends below 850-hPa. NeuralGCM does not capture near-surface warming over ocean and land, but does capture the profile structure over ocean (Fig.~\ref{Fig. 2}b, c). ACE2 also does not capture near-surface warming trends over the ocean (Fig.~\ref{Fig. 2}b), but does capture the near-surface trend over land (Fig.~\ref{Fig. 2}c). The reduced near-surface temperature trend around 925-hPa (Fig.~\ref{Fig. 2}b) over the ocean seems to be related to a cooling trend in the Eastern tropical Pacific that is stronger in the AI models than in the physics-based models (See Fig. S6).  The near-surface difference between ACE2 and NeuralGCM structure does not seem to be related to their training periods (See Fig. S5) and may be related to their different vertical resolutions.  Note that ACE2 only outputs two hybrid-sigma levels below 850-hPa to interpolate from. 

Previous studies have linked the excessive upper-tropospheric warming in physics-based models to limitations in their convection parameterizations. Here, both ACE2 and NeuralGCM substantially reduce this bias, suggesting that by learning the short-term weather trajectories they implicitly learned representations of subgrid-scale tendencies, including for moist convection, that improve their skill.

In the Northern extratropics, both AI models perform better than physics-based models in capturing the vertical temperature trends (Fig.~\ref{Fig. 2}d-f).  However, ACE2 performs best in reproducing the observed vertical temperature trend throughout the atmosphere (Fig.~\ref{Fig. 2}e). NeuralGCM struggles above 850-hPa over ocean (Fig.~\ref{Fig. 2}e) and above 700-hPa over land (Fig.~\ref{Fig. 2}f).  These findings are consistent regardless of their testing and training periods near the surface and aloft (See Fig. S5).  This difference between NeuralGCM and ACE2 is again surprising given that NeuralGCM has a dynamical core that one might expect would lead to more skill in capturing free tropospheric temperature.

\begin{figure}
\includegraphics{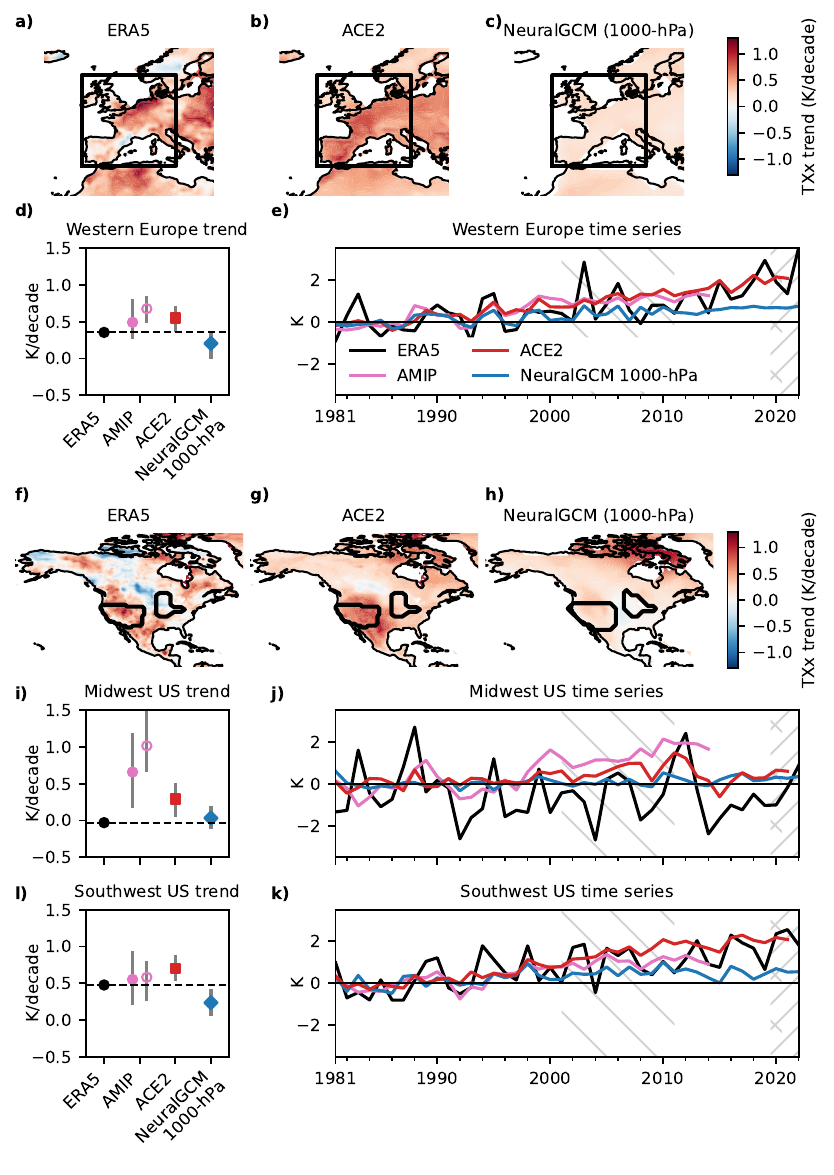}
\caption{2-meter TXx trends for ERA5, physics-based models, and ACE2, and 1000-hPa TXx trends for NeuralGCM (2-meter data are not available).  First row: Spatial TXx trends over Western Europe.  Second row: (d) Model ensemble spread of TXx trend averaged over Western Europe region outlined in black in the first row (1981-2014). Open circle represents CAM6. Grey vertical line represents 5-95\% spread. (e) Ensemble mean time series of latitude-weighted average TXx for Western Europe normalized by 1981-1990 average. Third row: As in first row, but for Midwest US. Fourth row: As in second row, but for Midwest US. The t-test did not show statistically significant tests for ERA5 and NeuralGCM. Last row: As in second row, but for Southwest US.}
\label{Fig. 3}
\end{figure}

\subsection{Regional Trends in Heat Extremes}

Previous work reported that physics-based coupled climate models struggle to capture regional trends in heat extremes over the Midwest US and Western Europe \cite{Singh2023}. Here, we examine trends in the annual maximum of daily maximum temperature ($\mathrm{TXx}$) in three regions (Fig.~\ref{Fig. 3}a-c, f-h) and find that the physics-based atmosphere-land (AMIP) ensemble captures the increase in heat extremes over Western Europe (Fig.~\ref{Fig. 3}d, e).  Over the Midwest US, physics-based models do not capture the ERA5 trend, overestimating heat extremes (Fig.~\ref{Fig. 3}i, j). 

Over the North American continent, the largest trend in heat extremes is in the US Southwest. The physics-based model ensemble captures the trend of heat extremes over the Southwest US (Fig.~\ref{Fig. 3}l, k). The CAM6 ensemble agrees with the AMIP ensemble over the Southwest US (Fig.~\ref{Fig. 3}l) but overestimates the trend in heat extremes over Western Europe and the Midwest US (Fig.~\ref{Fig. 3}d, i).

ACE2 and NeuralGCM show varying performance in capturing trends in heat extremes. NeuralGCM captures the trend over Western Europe and the Midwest US (Fig.~\ref{Fig. 3}d, i); however, it underestimates the trend over the Southwest US (Fig.~\ref{Fig. 3}l).  ACE2 shows similar performance to NeuralGCM over Western Europe and the Southwest US (Fig.~\ref{Fig. 3}d, l), however, it does not capture the trend over the Midwest US (Fig.~\ref{Fig. 3}i).  Generally, ACE2 shows larger trends in heat extremes (Fig.~\ref{Fig. 3}a, b, f, g) as compared to NeuralGCM, whose trends are closer to zero (Fig.~\ref{Fig. 3}a, c, f, h). 

The interannual variability of the ensemble mean time series in heat extremes in all models is smaller than the interannual variability in ERA5, which is just one realization of internal variability. The reduced interannual variability in the ensemble mean over time compared to ERA5 (Fig.~\ref{Fig. 3}e, j, k) seems to be related to ensemble averaging (see Fig. S7).  Both AI models show smaller variance over the Midwest US compared to ERA5, and NeuralGCM shows smaller variance over Western Europe (see Fig. S8).  The fact that AI models do not consistently capture heat extremes over land may indicate the need for more robust land representation in AI models to capture extreme temperatures.

\begin{figure}
\includegraphics{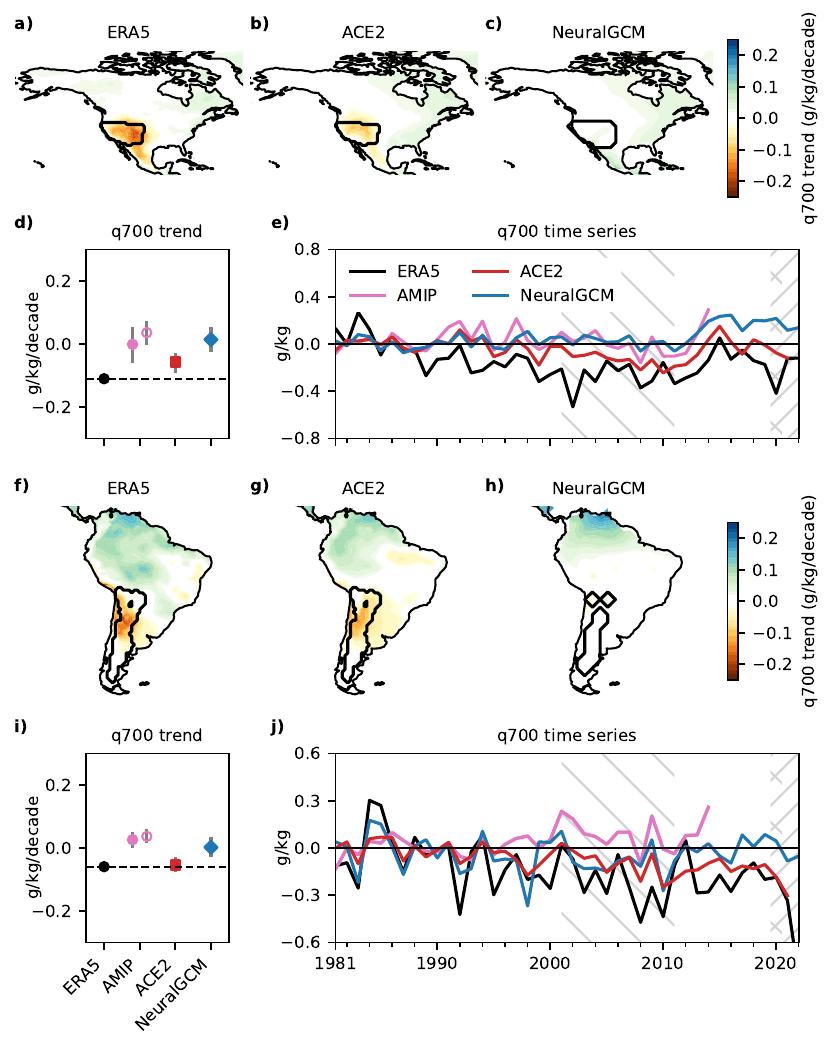}
\caption{700-hPa Specific humidity (q) trends by region. First row: Spatial specific humidity trend for ERA5, ACE2, and NeuralGCM for Southwest US.  Second row: (d) Model ensemble spread of humidity trend averaged over Southwest US region outlined in black (1981-2014). Open circle represents the CAM6 ensemble. Grey vertical line represents 5-95\% ensemble spread. The t-test did not show statistically significant trends for physics-based models or NeuralGCM.  (e) Ensemble mean annual time series of latitude-weighted average specific humidity for Southwest US normalized by 1981-1990 average. Backslash hatching represents the ACE2 testing period and forward slash hatching represents the NeuralGCM testing.  Third row: As in first row, but for South America. Last row: As in second row, but for South America. The t-test did not show statistically significant trends for physics-based models or NeuralGCM.}
\label{Fig. 4}
\end{figure}

\subsection{Regional Drying Trends}

Previous work reported drying in arid and semi-arid regions that was not captured by physics-based models \cite{Simpson2023, DryingRegionsDynamic}. Here we examine drying trends in two arid and semi-arid regions, the Southwest US (Fig.~\ref{Fig. 4}a-c) and Southern South America (Fig.~\ref{Fig. 4}f-h).  Consistent with previous work, the physics-based AMIP ensemble examined here does not capture the observed near-surface drying (700-hPa specific humidity trend due to regional topography) trend in the Southwest US and South America (Fig.~\ref{Fig. 4}d, e, i, j). The CAM6 ensemble does not capture the observed drying, consistent with the AMIP ensemble (Fig.~\ref{Fig. 4}d, i). 

The ACE2 and NeuralGCM ensembles also struggle to simulate drying trends over the Southwest US and South America (Fig.~\ref{Fig. 4}d, i).  NeuralGCM does not simulate drying trends over Southwest US (Fig.~\ref{Fig. 4}a, c, d) or South America (Fig.~\ref{Fig. 4}f, h, i). ACE2 shows a drying trend over the Southwest US closest to ERA5 (Fig.~\ref{Fig. 4}a, b, d) and captures the drying trend over South America (Fig.~\ref{Fig. 4}f, g, i). This performance does not depend on the testing and training period of these models (Fig.~\ref{Fig. 4}e, j).  Specific humidity trends for ERA5, physics-based models, and ACE2 at the 2-meter reference height agree with these findings (See Fig. S9). It may be that the lack of land features, such as soil moisture, is impacting the AI models' performance on near-surface drying trends.

\section{Conclusions \& Discussion}
    
\subsection{Conclusions}
In this study, we benchmark regional thermodynamic trends in an AI emulator, ACE2-ERA5, and a hybrid model, NeuralGCM, against ERA5 and physics-based models. Both AI models are trained on the weather task: capturing the short-term evolution of ERA5 trajectories, conditioned on prescribed SST and SIC (Eq.~(\ref{eq:loss})). The regional trend benchmarks quantifies their skill on a climate task. We focused specifically on regional trends over the satellite era that have been the focus of recent literature on physics-based models \cite{Singh2023, Simpson2023}. These include successes such as Arctic Amplification and land warming more than ocean, along with discrepancies, such as tropical upper-tropospheric warming, heat extremes, and regional drying.  

Overall, we find NeuralGCM performs as well as or better than physics-based models. It performs well in capturing regional temperature trends such as Arctic Amplification, but shows varying performance over other latitudinal regions (Fig.~\ref{Fig. 1}). For vertical tropical temperature trends, NeuralGCM shows good performance in the tropics but poor performance in the Northern Hemisphere extratropics. NeuralGCM captures the structure of near-surface warming over ocean, but not over land (Fig.~\ref{Fig. 2}a-c). In the Northern midlatitudes, NeuralGCM shows less warming bias than physics-based models (Fig.~\ref{Fig. 2}d-f). NeuralGCM is also able to capture the trend in heat extremes over Western Europe and the Midwest US, but does capture trends in heat extremes over the Southwest US (Fig.~\ref{Fig. 3}). NeuralGCM is unable to capture regional drying trends over the Southwest US and South America (Fig.~\ref{Fig. 4}), consistent with physics-based models.

The fully data-driven ACE2 model performs as well as other models for Arctic warming (Fig.~\ref{Fig. 1}), the vertical structure of warming in the tropics (Fig.~\ref{Fig. 2}a), and regional trends in heat extremes (Fig.~\ref{Fig. 3}). It does not capture the structure of near-surface warming over the ocean but better captures the profile over land in the tropics (Fig.~\ref{Fig. 2}b,c). Interestingly, we find ACE2 outperforms all other models in capturing the vertical structure of extratopical warming trends (Fig.~\ref{Fig. 2}) and regional drying in the US Southwest and South America (Fig.~\ref{Fig. 4}).

\subsection{Discussion}

Overall, the AI models benchmarked here show promising skill in capturing regional thermodynamic trends, a climate task, when they are only trained on the weather task \cite{watson-parrisMachineLearningWeather2021}. They perform as well as or better than physics-based models. Their improved performance relative to physics-based models, especially for upper-tropospheric warming, may be related to learning from ERA5 data. In particular, NeuralGCM replaces parameterized tendencies with a neural network trained on ERA5, while ACE2 is completely data-driven. Learning from data may therefore mitigate biases introduced by traditional subgrid-scale parameterizations, highlighting the potential of AI to improve subgrid-scale representations in physics-based models \cite{bracco2025machine, Tapio_parameterization, Gentine_parameterization}.  

As one moves from larger (e.g., Arctic Amplification) to smaller (e.g., US Southwest) regional scales, the skill of NeuralGCM and ACE2 varies more, which may result from their global loss function and best-model-selection procedures (based on checkpoints, random seeds, and hyperparameters), which do not prioritize regional biases. For example, errors could be minimized globally but redistributed differently across regions. Furthermore, spectral bias/blurring \cite{chattopadhyay2023challenges,Bonavita_2024} and data imbalance \cite{Sun_2025,sun2025predicting, Materia_ai_for_climate_extremes}, could explain the poor performance of AI models for heat extremes in some regions. The models show reduced TXx variability over some of the regions considered in this study (See Fig. S8).  In addition, NeuralGCM is unable to capture drying trends, and ACE2 is unable to capture the trend over the Southwest US. AI models performance on both extreme temperatures and drying trends highlights the fact that they do not include explicit land representation.  Though further testing is needed to prove causation, the inclusion of land features in training could improve near-surface skill over land.

Our results show that ACE2 exhibits comparable and at times, better performance compared to NeuralGCM suggesting a dynamical core does not necessarily lead to increased skill in capturing the regional thermodynamic trends examined here. Most notably, ACE2 captures trends of warming in the Northern and Southern midlatitude troposphere better than NeuralGCM.  ACE2 might perform better because it does not have the complexities of integrating the dynamical core and neural network-based parameterization.  For example, NeuralGCM's dynamical core drifts without a surface pressure correction \cite{ngcm-imerg}, so, in addition to learning the subgrid-scale tendencies, the neural network would need to compensate for biases in the dynamical core, which could cause other biases.  Another reason for ACE2's sometimes better performance against NeuralGCM could be that it includes carbon dioxide forcing, perhaps giving it more information to emulate climate signals. 

It should be stressed that the AI models examined here were benchmarked for a historical, i.e., in-distribution, climate task involving only the atmosphere-land processes (forced with SST and SIC). They are not coupled climate models that include greenhouse gases, aerosols, cloud processes, and an ocean component (such models have just started to emerge \cite{SamudrACE, DLESym}).  Thus, the benchmarks examined here do not test out-of-distribution climate scenarios such as +2K to 4K SST \cite{+2k_emulation, camulator, PNNL_Hierarchical}.

Our results highlight the value of benchmarking AI models alongside physics-based models when examining skill in capturing satellite-era regional thermodynamic trends. The AI models show great promise with their computational efficiency and ability to learn the weather task and apply it to the climate task. Future work should focus on the representation of land-atmosphere interactions by training on land variables such as soil moisture, which has the potential to improve the AI models' skill in capturing heat extremes \cite{ngcm_heatwaves_land_feedbacks} and near-surface drying trends.

%
%

\section*{Open Research Section}
ERA5 is publicly available from the Copernicus Climate Change Service (C3S) Climate Data Store (CDS) at \url{https://cds.climate.copernicus.eu/cdsapp#!/dataset/}. AMIP data was collected from the Earth System Grid Federation (ESGF) at \url{https://esgf-node.llnl.gov/projects/cmip6/}.json.  NeuralGCM checkpoints and code can be found at \url{https://github.com/neuralgcm/neuralgcm}. ACE2 checkpoints and code can be found at \url{https://huggingface.co/allenai/ACE2-ERA5} and \url{https://github.com/ai2cm/ace}.  Codes and data for computing diagnostics and recreating figures can be found at \cite{Data-codes}.

\section*{Acknowledgments}
K.R. and P.H. are supported by NSF AGS-2531264 and Schmidt Sciences, LLC. H.P is supported by NSF OAC-2544065. I.B. and T.A.S. are supported by National Oceanic and Atmospheric Administration award NA23OAR4310597.  T.A.S also acknowledges support from the Simons Foundation (grant SFI-MPS-T-MPS-00007353). Computational resources by NSF ACCESS (ATM170020), NCAR’s CISL (URIC0009), and the University of Chicago Research Computing Center.

%
\bibliographystyle{unsrt}  
\bibliography{agusample}

\begin{thebibliography}{10}

\bibitem{IPCC2021}
IPCC.
\newblock {\em “Summary for Policymakers”}, pages 33\--144.
\newblock Cambridge University Press, Cambridge, United Kingdom and New York, NY, USA, 2021.

\bibitem{Wallace_Joshi2018}
C~J Wallace and M~Joshi.
\newblock Comparison of land-ocean warming ratios in updated observed records and cmip5 climate models.
\newblock {\em Environmental Research Letters}, 12, 2018.

\bibitem{Hahn2021}
L.~C. Hahn, K.~C. Armour, M.~D. Zelinka, C.~M. Bitz, and A.~Donohoe.
\newblock Contributions to polar amplification in cmip5 and cmip6 models.
\newblock {\em Frontiers in Earth Science}, Volume 9 - 2021, 2021.

\bibitem{Manabe&Wetherald}
Syukuro Manabe and Richard~T. Wetherald.
\newblock On the distribution of climate change resulting from an increase in co2 content of the atmosphere.
\newblock {\em Journal of Atmospheric Sciences}, 37(1):99 -- 118, 1980.

\bibitem{Shaw2024}
Tiffany~A. Shaw, Paola~A. Arias, Mat Collins, Dim Coumou, Arona Diedhiou, Chaim~I. Garfinkel, Shipra Jain, Mathew~Koll Roxy, Marlene Kretschmer, L.~Ruby Leung, Sugata Narsey, Olivia Martius, Richard Seager, Theodore~G. Shepherd, Anna~A. Sörensson, Tannecia Stephenson, Michael Taylor, and Lin Wang.
\newblock Regional climate change: consensus, discrepancies, and ways forward.
\newblock {\em Frontiers in Climate}, Volume 6 - 2024, 2024.

\bibitem{Vert_profile_CMIP6}
Dann~M Mitchell, Y~T~Eunice Lo, William J~M Seviour, Leopold Haimberger, and Lorenzo~M Polvani.
\newblock The vertical profile of recent tropical temperature trends: Persistent model biases in the context of internal variability.
\newblock {\em Environmental Research Letters}, 15(10), 2020.

\bibitem{Singh2023}
Jitendra Singh, Sebastian Sippel, and Erich~M. Fischer.
\newblock Circulation dampened heat extremes intensification over the midwest usa and amplified over western europe.
\newblock {\em Communications Earth \& Environment}, 4(432), 2023.

\bibitem{Simpson2023}
Isla~R. Simpson, Karen~A. McKinnon, Daniel Kennedy, David~M. Lawrence, Flavio Lehner, and Richard Seager.
\newblock Observed humidity trends in dry regions contradict climate models.
\newblock {\em Proceedings of the National Academy of Sciences}, 121(1):e2302480120, 2024.

\bibitem{shaw2025}
Tiffany~A. Shaw and Bjorn Stevens.
\newblock The other climate crisis.
\newblock {\em Nature}, 639:877--887, 2025.

\bibitem{forecastnet}
Jaideep Pathak, Shashank Subramanian, Peter Harrington, Sanjeev Raja, Ashesh Chattopadhyay, Morteza Mardani, Thorsten Kurth, David Hall, Zongyi Li, Kamyar Azizzadenesheli, Pedram Hassanzadeh, Karthik Kashinath, and Animashree Anandkumar.
\newblock Fourcastnet: A global data-driven high-resolution weather model using adaptive fourier neural operators, 2022.

\bibitem{pangu-weather}
Kaifeng Bi, Lingxi Xie, Hengheng Zhang, Xin Chen, Xiaotao Gu, and Qi~Tian.
\newblock Accurate medium-range global weather forecasting with 3d neural networks.
\newblock {\em Nature}, 619:1--6, 07 2023.

\bibitem{graphcast}
Remi Lam, Alvaro Sanchez-Gonzalez, Matthew Willson, Peter Wirnsberger, Meire Fortunato, Ferran Alet, Suman Ravuri, Timo Ewalds, Zach Eaton-Rosen, Weihua Hu, Alexander Merose, Stephan Hoyer, George Holland, Oriol Vinyals, Jacklynn Stott, Alexander Pritzel, Shakir Mohamed, and Peter Battaglia.
\newblock Learning skillful medium-range global weather forecasting.
\newblock {\em Science}, 382(6677):1416--1421, 2023.

\bibitem{Ben_Bouall_gue_2024}
Zied Ben~Bouallègue, Mariana C.~A. Clare, Linus Magnusson, Estibaliz Gascón, Michael Maier-Gerber, Martin Janoušek, Mark Rodwell, Florian Pinault, Jesper~S. Dramsch, Simon T.~K. Lang, Baudouin Raoult, Florence Rabier, Matthieu Chevallier, Irina Sandu, Peter Dueben, Matthew Chantry, and Florian Pappenberger.
\newblock The rise of data-driven weather forecasting: A first statistical assessment of machine learning–based weather forecasts in an operational-like context.
\newblock {\em Bulletin of the American Meteorological Society}, 105(6):E864–E883, June 2024.

\bibitem{ace2}
Oliver Watt-Meyer, Brian Henn, Jeremy McGibbon, Spencer~K. Clark, Anna Kwa, W.~Andre Perkins, Elynn Wu, Lucas Harris, and Christopher~S. Bretherton.
\newblock Ace2: Accurately learning subseasonal to decadal atmospheric variability and forced responses.
\newblock {\em npj Climate and Atmospheric Science}, 8(205 (2025)), 2025.

\bibitem{ngcm}
Dmitrii Kochkov, Janni Yuval, Ian Langmore, Peter Norgaard, Jamie Smith, Griffin Mooers, Milan Klöwer, James Lottes, Stephan Rasp, Peter Düben, Sam Hatfield, Peter Battaglia, Alvaro Sanchez-Gonzalez, Matthew Willson, Michael~P. Brenner, and Stephan Hoyer.
\newblock Neural general circulation models for weather and climate.
\newblock {\em Nature}, 632(8027):1060–1066, 2024.

\bibitem{camulator}
William~E. Chapman, John~S. Schreck, Yingkai Sha, David John~Gagne II, Dhamma Kimpara, Laure Zanna, Kirsten~J. Mayer, and Judith Berner.
\newblock Camulator: Fast emulation of the community atmosphere model, 2025.

\bibitem{DLESym}
Nathaniel Cresswell-Clay, Bowen Liu, Dale Durran, Zihui Liu, Zachary~I. Espinosa, Raul Moreno, and Matthias Karlbauer.
\newblock A deep learning earth system model for efficient simulation of the observed climate, 2025.

\bibitem{SamudrACE}
James P.~C. Duncan, Elynn Wu, Surya Dheeshjith, Adam Subel, Troy Arcomano, Spencer~K. Clark, Brian Henn, Anna Kwa, Jeremy McGibbon, W.~Andre Perkins, William Gregory, Carlos Fernandez-Granda, Julius Busecke, Oliver Watt-Meyer, William~J. Hurlin, Alistair Adcroft, Laure Zanna, and Christopher Bretherton.
\newblock Samudrace: Fast and accurate coupled climate modeling with 3d ocean and atmosphere emulators, 2025.

\bibitem{LUCIE}
Haiwen Guan, Troy Arcomano, Ashesh Chattopadhyay, and Romit Maulik.
\newblock Lucie-3d: A three-dimensional climate emulator for forced responses, 2025.

\bibitem{Mu-ting2025}
{Mu-ting} Chien, Elizabeth Barnes, and Eric~D Maloney.
\newblock Modulation of tropical cyclogenesis on subseasonal-to-interannual timescales in the deep-learning climate emulator ace2.
\newblock {\em Machine Learning: Earth}, 2025.

\bibitem{PNNL_Hierarchical}
Ziming Chen, {L. Ruby} Leung, Wenyu Zhou, Jian Lu, {Sandro W.} Lubis, Ye~Liu, {Chuan-Chieh} Chang, Bryce~E Harrop, Ya~Wang, Gan Zhang, and Yun Qian.
\newblock Hierarchal testing of a hybrid machine learning-physics global atmosphere model.
\newblock {\em ESS Open Archive}, 2025.

\bibitem{+2k_emulation}
Bosong Zhang and Timothy~M. Merlis.
\newblock The equilibrium response of atmospheric machine-learning models to uniform sea surface temperature warming, 2025.

\bibitem{Benchmark_variability}
Ian Baxter, Hamid~A. Pahlavan, Pedram Hassanzadeh, Katharine Rucker, and Tiffany~A. Shaw.
\newblock Benchmarking atmospheric circulation variability in an ai emulator, ace2, and a hybrid model, neuralgcm, 2025.

\bibitem{ngcm_heatwaves_land_feedbacks}
Shiheng Duan, Jishi Zhang, Céline Bonfils, and Giuliana Pallotta.
\newblock Testing neuralgcm's capability to simulate future heatwaves based on the 2021 pacific northwest heatwave event, 2025.

\bibitem{chattopadhyay2023challenges}
Ashesh Chattopadhyay, Y~Qiang Sun, and Pedram Hassanzadeh.
\newblock Challenges of learning multi-scale dynamics with ai weather models: Implications for stability and one solution.
\newblock {\em arXiv e-prints}, pages arXiv--2304, 2023.

\bibitem{Bonavita_2024}
Massimo Bonavita.
\newblock On some limitations of current machine learning weather prediction models.
\newblock {\em Geophysical Research Letters}, 51(12):e2023GL107377, 2024.
\newblock e2023GL107377 2023GL107377.

\bibitem{Lai_2025}
Ching-Yao Lai, Pedram Hassanzadeh, Aditi Sheshadri, Maike Sonnewald, Raffaele Ferrari, and Venkatramani Balaji.
\newblock Machine learning for climate physics and simulations.
\newblock {\em Annual Review of Condensed Matter Physics}, 16(Volume 16, 2025):343--365, 2025.

\bibitem{Sun_2025}
Y.~Qiang Sun, Pedram Hassanzadeh, Mohsen Zand, Ashesh Chattopadhyay, Jonathan Weare, and Dorian~S. Abbot.
\newblock Can ai weather models predict out-of-distribution gray swan tropical cyclones?
\newblock {\em Proceedings of the National Academy of Sciences}, 122(21):e2420914122, 2025.

\bibitem{sun2025predicting}
Y~Qiang Sun, Pedram Hassanzadeh, Tiffany Shaw, and Hamid~A Pahlavan.
\newblock Predicting beyond training data via extrapolation versus translocation: Ai weather models and dubai's unprecedented 2024 rainfall.
\newblock {\em arXiv preprint arXiv:2505.10241}, 2025.

\bibitem{Materia_ai_for_climate_extremes}
Stefano Materia, Lluís~Palma García, Chiem van Straaten, Sungmin O, Antonios Mamalakis, Leone Cavicchia, Dim Coumou, Paolo de~Luca, Marlene Kretschmer, and Markus Donat.
\newblock Artificial intelligence for climate prediction of extremes: State of the art, challenges, and future perspectives.
\newblock {\em WIREs Climate Change}, 15(6):e914, 2024.

\bibitem{gates1992}
W.~Lawrence Gates.
\newblock {AMIP}: The atmospheric model intercomparison project.
\newblock {\em Bull. Amer. Met. Soc.}, 73(8):1962--1970, 1992.

\bibitem{eyring_CMIP6}
V.~Eyring, S.~Bony, G.~A. Meehl, C.~A. Senior, B.~Stevens, R.~J. Stouffer, and K.~E. Taylor.
\newblock Overview of the coupled model intercomparison project phase 6 (cmip6) experimental design and organization.
\newblock {\em Geoscientific Model Development}, 9(5):1937--1958, 2016.

\bibitem{ngcm-imerg}
Janni Yuval, Ian Langmore, Dmitrii Kochkov, and Stephan Hoyer.
\newblock Neural general circulation models optimized to predict satellite-based precipitation observations, 2024.

\bibitem{Po-Chedley2022}
Stephen Po-Chedley, John~T. Fasullo, Nicholas Siler, Zachary~M. Labe, Elizabeth~A. Barnes, Céline J.~W. Bonfils, and Benjamin~D. Santer.
\newblock Internal variability and forcing influence model–satellite differences in the rate of tropical tropospheric warming.
\newblock {\em Proceedings of the National Academy of Sciences}, 119(47):e2209431119, 2022.

\bibitem{Fueglistaler2025}
S.~Fueglistaler, C.~Radley, and I.~M. Held.
\newblock The distribution of precipitation and the spread in tropical upper tropospheric temperature trends in cmip5/amip simulations.
\newblock {\em Geophysical Research Letters}, 42(14):6000--6007, 2025.

\bibitem{Mitchell_2013}
D.~M. Mitchell, P.~W. Thorne, P.~A. Stott, and L.~J. Gray.
\newblock Revisiting the controversial issue of tropical tropospheric temperature trends.
\newblock {\em Geophysical Research Letters}, 40(11):2801--2806, 2013.

\bibitem{DryingRegionsDynamic}
Robert Doane-Solomon, Tim Woollings, and Isla~R. Simpson.
\newblock Dynamic {{Contributions}} to {{Recent Observed Wintertime Precipitation Trends}} in {{Mediterranean-Type Climate Regions}}.
\newblock 52(12):e2024GL114258, 2025.

\bibitem{watson-parrisMachineLearningWeather2021}
D.~{Watson-Parris}.
\newblock Machine learning for weather and climate are worlds apart.
\newblock {\em Philosophical Transactions of the Royal Society A: Mathematical, Physical and Engineering Sciences}, 379(2194):20200098, February 2021.

\bibitem{bracco2025machine}
Annalisa Bracco, Julien Brajard, Henk~A Dijkstra, Pedram Hassanzadeh, Christian Lessig, and Claire Monteleoni.
\newblock Machine learning for the physics of climate.
\newblock {\em Nature Reviews Physics}, 7(1):6--20, 2025.

\bibitem{Tapio_parameterization}
T.~Schneider, L.~R. Leung, and R.~C.~J. Wills.
\newblock Opinion: Optimizing climate models with process knowledge, resolution, and artificial intelligence.
\newblock {\em Atmospheric Chemistry and Physics}, 24(12):7041--7062, 2024.

\bibitem{Gentine_parameterization}
Yongquan Qu, Mohamed~Aziz Bhouri, and Pierre Gentine.
\newblock Joint parameter and parameterization inference with uncertainty quantification through differentiable programming, 2024.

\bibitem{Data-codes}
Katharine Rucker.
\newblock Codes and data, 2025.

\end{thebibliography}


\clearpage
\appendix
\section*{Supplementary Information (SI)}

\renewcommand{\thefigure}{S\arabic{figure}} 
\setcounter{figure}{0}

\pagebreak 

\begin{table}
    \centering
    \caption{AMIP models used in this study}
    \begin{tabular}{c c c c}
    \hline
    \multicolumn{4}{ c }{AMIP models} \\
    \hline
    Model Name & Realizations & Model Name & Realizations \\
    \hline 
    ACCESS-CM2 & 7 & GISS-E2-2-G & 1 \\
    ACCESS-ESM1-5 & 3 & HadGEM3-GC31-LL & 5 \\
    BCC-CSM2-MR & 3 & HadGEM3-GC31-MM & 4 \\
    BCC-ESM1 & 3 & IITM-ESM & 1 \\
    CAMS-CSM1-0 & 3 & INM-CM4-8 & 1 \\
    CESM2-FV2 & 3 & INM-CM5-0 & 1 \\
    CESM2-WACCM-FV2 & 1 & IPSL-CM6A-LR & 1 \\
    CESM2-WACCM & 3 & KACE-1-0-G & 1 \\
    CESM2 & 10 & KIOST-ESM & 1 \\
    CIESM & 3 & MIROC-ES2L & 3 \\
    CMCC-CM2-HR4 & 1 &  MIROC6 & 10 \\
    CMCC-CM2-SR5 & 1 & MPI-ESM-1-2-HAM & 1 \\
    CNRM-CM6-1-HR & 1 & MPI-ESM1-2-HR & 3 \\
    CNRM-CM6-1 & 1 & MPI-ESM1-2-LR & 3 \\
    CNRM-ESM2-1 & 1 & MRI-ESM2-0 & 3 \\
    CanESM5 & 7 & NESM3 & 5 \\
    E3SM-1-0 & 3 & NorESM2-LM & 2\\
    EC-Earth3-AerChem & 2 & SAM0-UNICON & 1\\
    EC-Earth3-CC & 1 & TaiESM1 & 1 \\
    EC-Earth3-Veg-LR & 1 & UKESM1-0-LL & 1 \\
    EC-Earth3-Veg & 1 \\
    EC-Earth3 & 1 \\
    FGOALS-f3-L & 3 \\
    FGOALS-g3 & 1 \\
    GFDL-AM4 & 1 \\
    GFDL-CM4 & 1 \\
    GFDL-ESM4 & 1 \\
    GISS-E2-1-G & 15 \\
    \end{tabular}
    \label{table: Models list}
\end{table}

\clearpage

\begin{figure}
    \centering
    \includegraphics{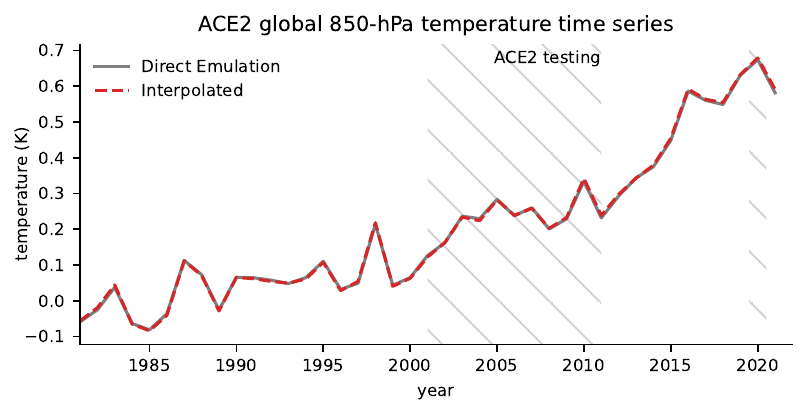}
    \caption{Comparison of normalized global temperature at 850-hPa from ACE2 direct output and at 850-hPa interpolated from ACE2's hybrid sigma levels.}
\end{figure}

\begin{figure}
    \centering
    \includegraphics{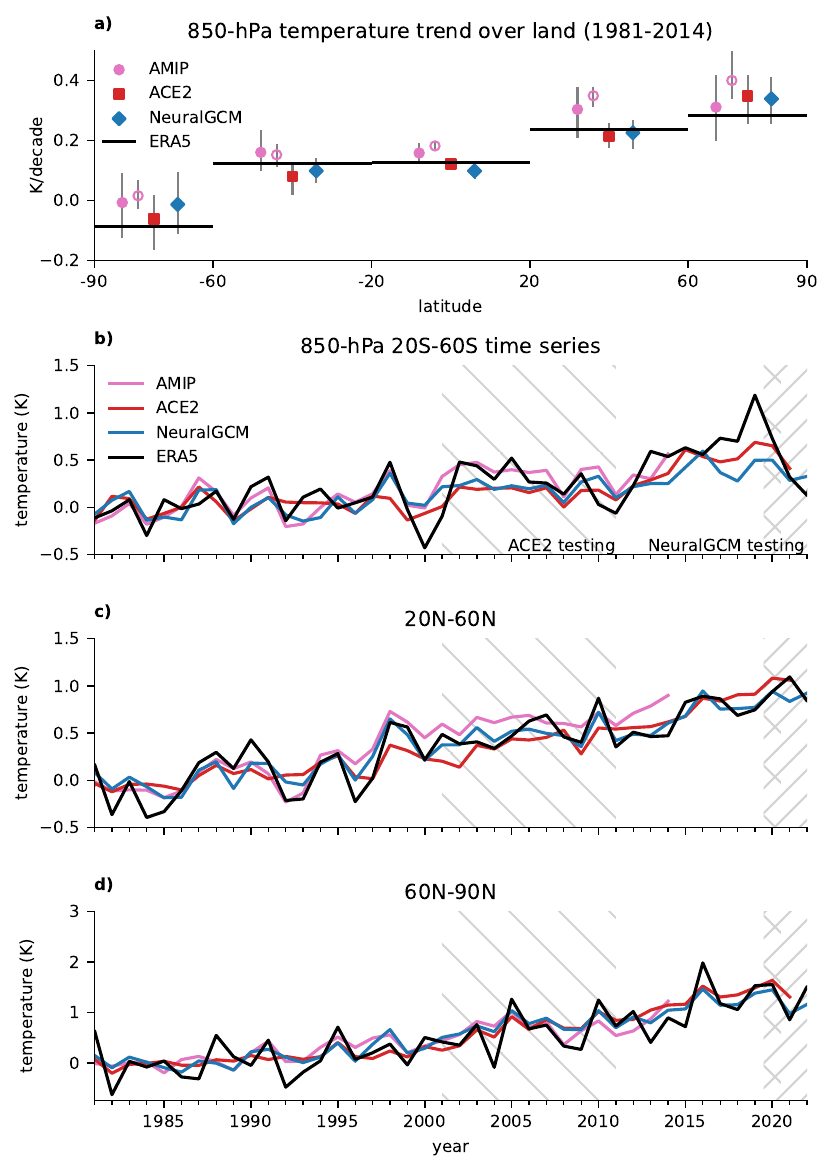}
    \caption{(a) Latitude-weighted zonal-mean 850-hPa temperature trend (1981-2014) in latitudinal sections for physics-based models, AI models and ERA5 over land.  Open pink circle represents the CAM6 ensemble. Grey vertical lines represent the 5-95\% ensemble spread. Black horizontal line is the ERA5 trend. (b) Ensemble mean annual time series of latitude-weighted average temperature over 20S-60S at 850-hPa normalized by 1981-1990 average. (c) As in (b) but for 20N-60N. (d) As is (b) but for 60N-90N.}
\end{figure}

\begin{figure}
    \centering
    \includegraphics{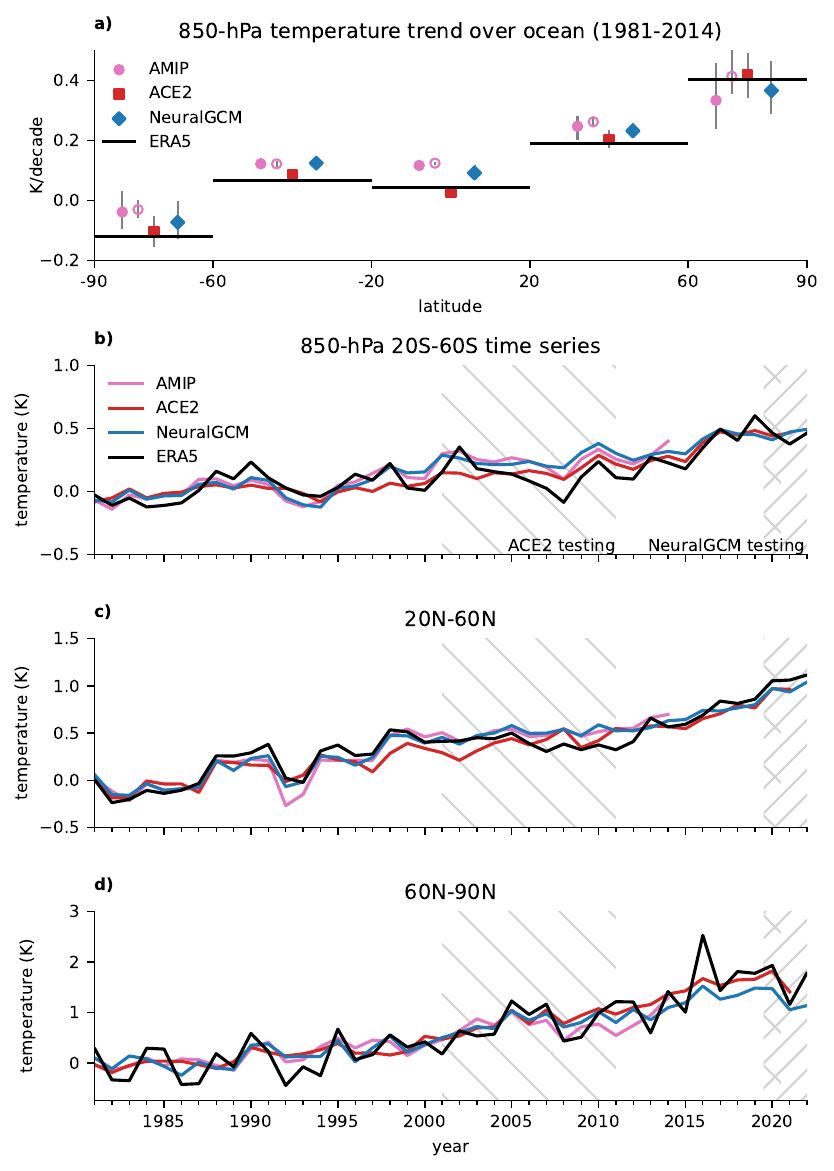}
    \caption{(a) Latitude-weighted zonal-mean 850-hPa temperature trend (1981-2014) in latitudinal sections for physics-based models, AI models, and ERA5 over ocean.  Open pink circle represents the CAM6 ensemble. Grey vertical lines represent the 5-95\% ensemble spread. Black horizontal line is the ERA5 trend. (b) Ensemble mean annual time series of latitude-weighted average temperature over 20S-60S at 850-hPa normalized by 1981-1990 average. (c) As in (b) but for 20N-60N. (d) As is (b) but for 60N-90N.}
\end{figure}

\begin{figure}
    \centering
    \includegraphics{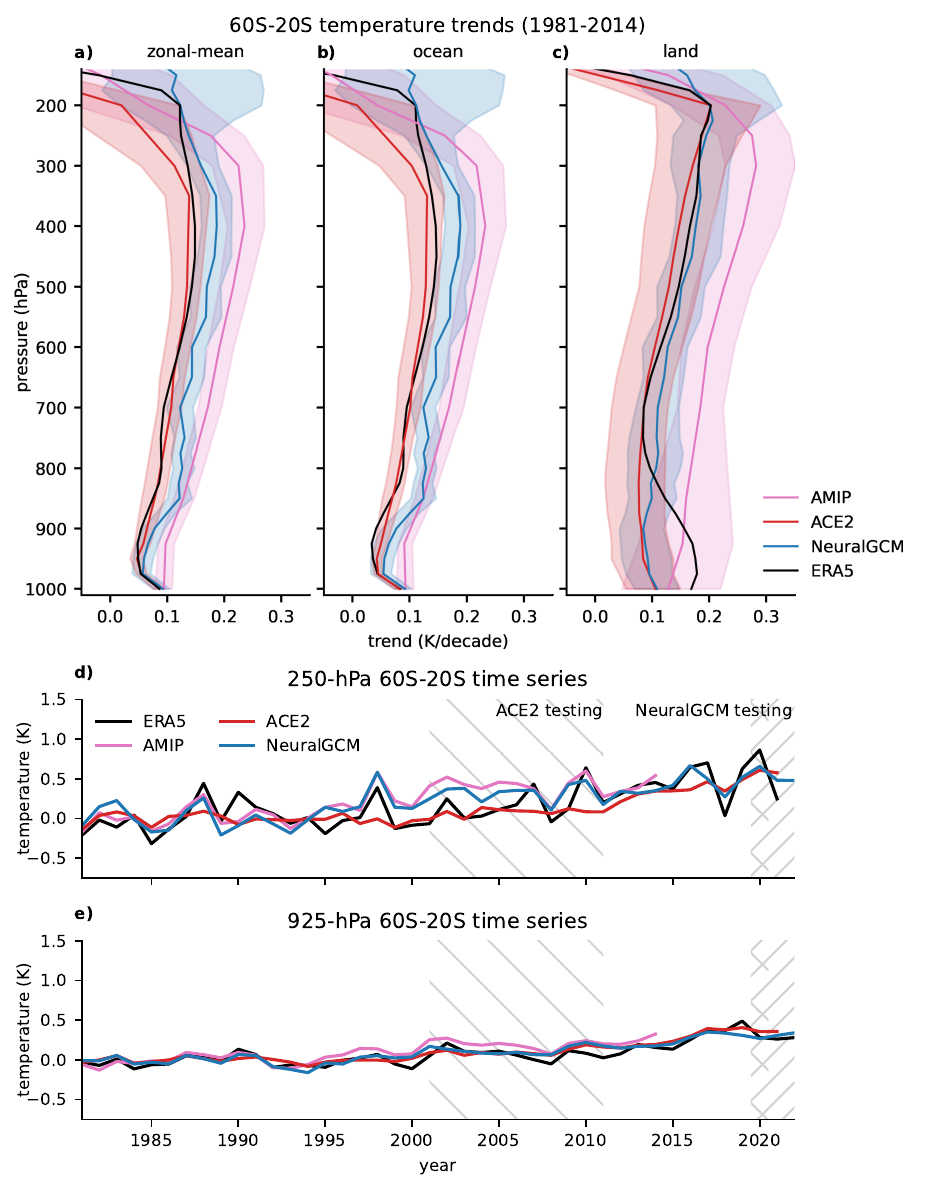}
    \caption{(a)-(c) Temporal and spatial average vertical profile of southern mid-latitude (20S-60S) temperature trends of physics-based models and AI models compared to ERA5.  Shading shows the ensemble spread 5-95\% (d) Annual time series of latitude-weighted average of ensemble mean at 250-hPa over 20S-60S normalized by 1981-1990 average of each model. Backslash hatching represent the ACE2 testing period and forward slash hatching represents the NGCM testing period.}
\end{figure}

\begin{figure}
    \centering
    \includegraphics{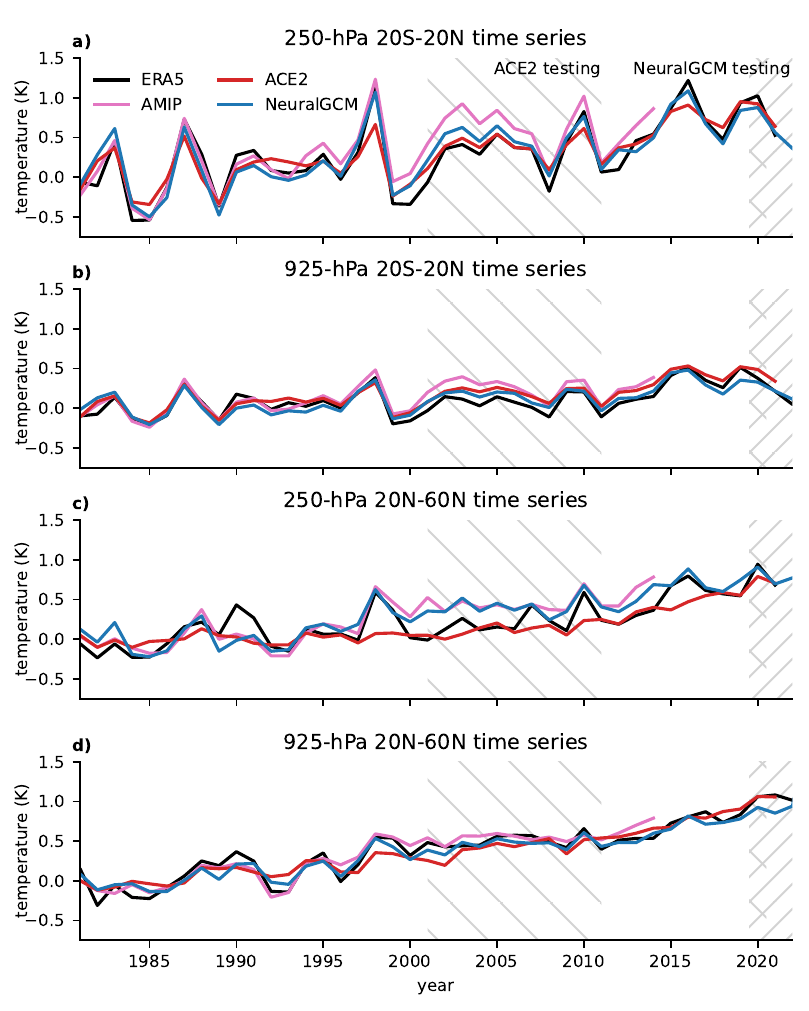}
    \caption{(a) Annual time series of latitude-weighted average of ensemble mean at 250-hPa over 20S-20N normalized by 1981-1990 average of each model. Backslash hatching represent the ACE2 testing period and forward slash hatching represents the NGCM testing period. (b) As in (a) but for 925-hPa. (c) As in (a) but 20N-60N. (d) As in (b) but for 20N-60N.}
\end{figure}

\begin{figure}
    \centering
    \includegraphics{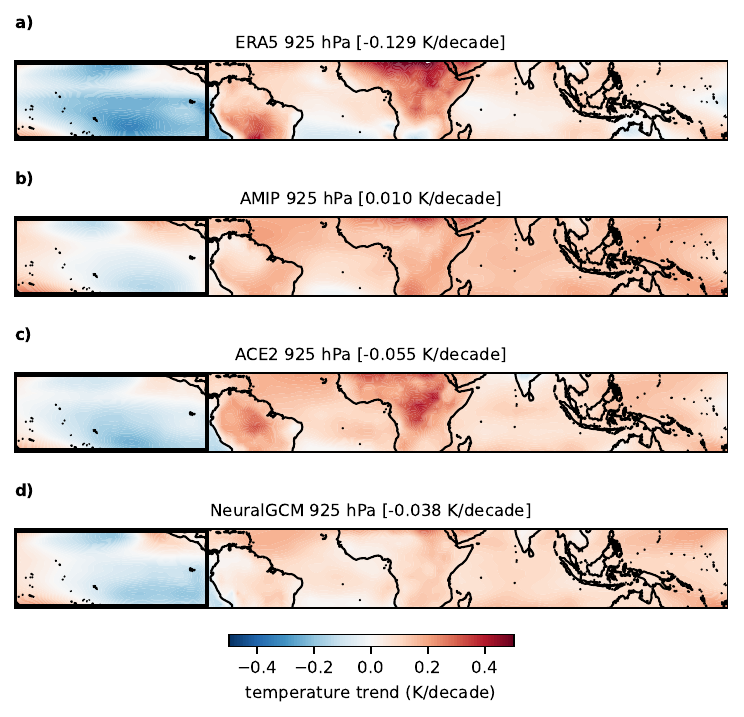}
    \caption{Spatial map of temperature trends at 925 hPa in ERA5, ensemble mean of physics-based models, ensemble mean of ACE2, and ensemble mean of NeuralGCM. Average trend over boxed region is indicated to the right of the title for each panel.}
\end{figure}

\begin{figure}
    \centering
    \includegraphics{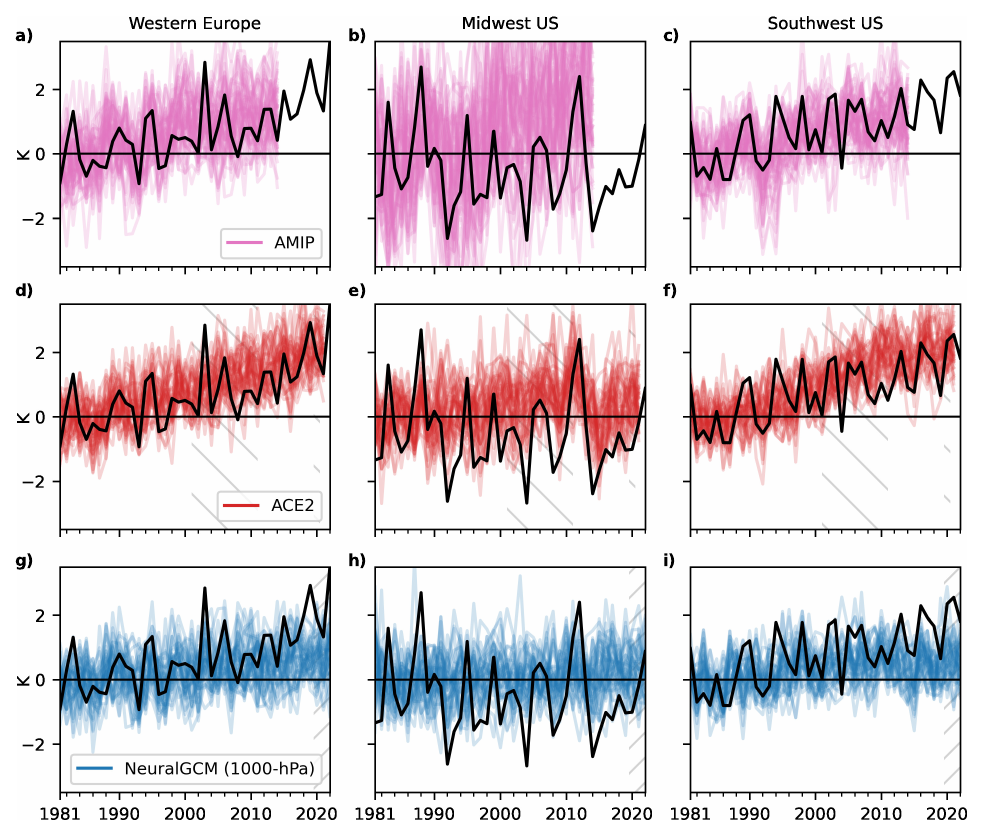}
    \caption{First row: AMIP ensemble time series of latitude-weighted average annual maximum 2-meter temperature for Western Europe (left), Midwest US (center), and Southwest US (right) normalized by 1981-1990 average. Second row: As in first row, but for ACE2. Backslash hatching represents the testing period. Third row: As in first row but for NeuralGCM at 1000-hPa. Forward slash hatching represents the testing period.}
\end{figure}

\begin{figure}
    \centering
    \includegraphics{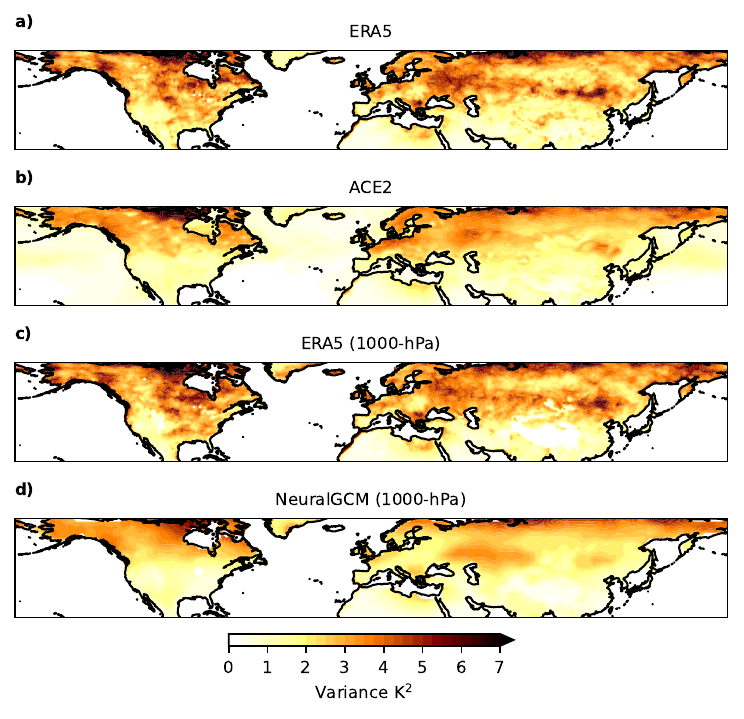}
    \caption{Detrended annual maximum of daily maximum temperature time-variance maps of (a) ERA5 2-meter, (b) ensemble mean ACE2 2-meter, (c) ERA5 1000-hPa, and (d) ensemble mean NeuralGCM 1000-hPa.}
\end{figure}

\begin{figure}
    \includegraphics{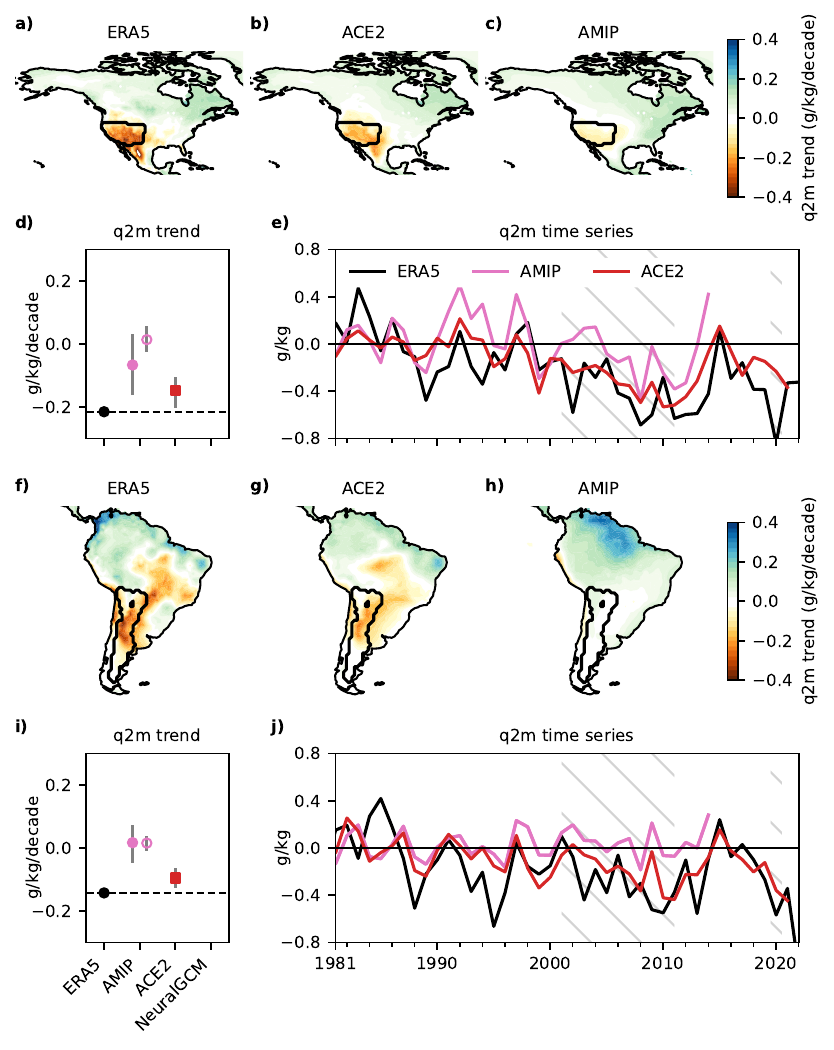}
    \caption{2-meter Specific humidity (q) trends by region. First row: Spatial specific humidity trend for ERA5, ACE2, and physics-based models for Southwest US.  Second row: (d) Model ensemble spread of humidity trend averaged over Southwest US region outlined in black (1981-2014). Open circle represents the CAM6 ensemble. Grey vertical line represent 5-95\% ensemble spread. (e) Ensemble mean annual time series of latitude-weighted average specific humidity for Southwest US normalized by 1981-1990 average. Backslash hatching represents the ACE2 testing period.  Third row: As in first row, but for South America. Last row: As in second row, but for South America.}
\end{figure}

\clearpage

\end{document}